# Enhancement of two-photon transition rate by selectively removing certain frequency comb teeth

Shuangyou Zhang, Wanpeng Yang, and Jianye Zhao[*]

*Department of Electronics, School of Electronics Engineering and Computer Sciences, Peking University, Beijing 100871, China*

We present experiments demonstrating an enhancement of resonant two-photon transition rate in $^{87}$Rb utilizing spectral phase manipulation of the excitation frequency comb. By selectively removing certain comb teeth, the resonant two-photon transition rate can be improved, and reach a factor of more than 1.8. The femtosecond pulse-train excitation of two-photon transition is investigated theoretically based on general multiphoton transitions and the results are compared with the experiments. The theory presented here gives a clear insight of physical mechanism of this quantum coherent control and indicates that it is simple, effective and universal for nonlinear interactions between frequency combs and matters.

There has been a great deal of interest in two-photon spectroscopy carried out in atomic vapors by continuous wave (CW) lasers or frequency combs [1-4]. Two-photon transitions (TPTs) have proved to be a powerful tool for several branches of science, e.g., atomic spectroscopy and time domain ultrafast dynamics [5, 6]. As a classical example, 1S-2S two-photon spectroscopy with a natural linewidth of only 1-Hz in hydrogen is essential in the determination of the Rydberg constant and electron-proton mass ratio, accurate tests of quantum electrodynamics (QED) [3, 5]. Another classical example of TPTs is Rubidium 5S-5D or 5S-7S TPT, because of its low cost, intrinsic high metrological performances and excellent frequency properties for optical communication bands [6-9], and relevant optical frequency standards have been demonstrated with a high stability on the order of $10^{-13}\tau^{-1}$ for integration time $\tau$ up to 1000 s [7, 10].

With the rapid developments in the past decades, frequency combs [11] have played an essential and critical role in the laser spectroscopy, not only as high-precision frequency rules [12, 13], connecting tremendous frequency difference between radio frequency and optical frequency, but also employed to directly induce atomic transitions for laser spectroscopy, marking the beginning of a new field of direct frequency comb spectroscopy (DFCS) [6, 14-17]. Two-photon spectroscopy based on DFCS can perform a same or even better precision than that based on CW lasers [8, 17]. Moreover, due to high peak power of utlrashort pulses, DFCS paves the way to precisely measure transitions in the vacuum ultraviolet (VUV) region of the optical spectrum — wavelengths between approximately 10 and 200 nm [18], therefore has drawn increasing attention in laser spectroscopy.

Quantum coherent control of TPTs in the interaction between frequency combs and matters has been a research hotspot for a few decades, which dedicates to enhance the desired state or cancel out the undesirable outcomes [4, 19-23]. Some of the pioneering work on quantum coherent control has been demonstrated, including varying the spectral phases [4, 19-21], tailoring the comb spectrum [20, 22] and introducing chirp on the combs [23]. In most previous works [4, 19-21], spatial light modulators (SLMs) were employed to manipulate the spectral phases of Ti: Sapphire femtosecond lasers to excite the thermal or trapped atoms for changing the TPT rate. This can be greatly changed because the negative and positive contributions were comparable due to the broad bandwidth of femtosecond lasers. In this letter, we propose a particularly simple and effective coherent quantum control to enhance resonant TPT rate for universal optical frequency combs. Different from the previous work, we selectively remove some certain comb teeth instead of part of continuous optical spectrum or SLMs. As a result, the resonant TPT rate can be effectively improved by a factor of 1.8.

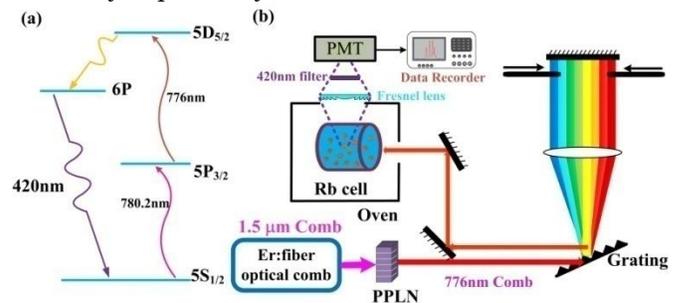

FIG.1 (color online). (a) Energy level diagram of the 87Rb. (b) Experimental setup. The components are the following: PPLN, periodically poled lithium niobate; PMT, photomultiplier tube.

When considering the TPTs in $^{87}$Rb between the 5S and the 5D states, the situation is more complicated since two types of TPTs are simultaneously happening, can both contribute to the total TPT intensity [24]. One is direct TPT (D-TPT), exited by a pair of comb modes together. The other is sequential TPT (S-TPT), correspond to the stepwise transitions via the intermediate state. Figure 1(a) is the simplified schematic representation of the energy levels of [87]Rb. Because the intermediate $5P_{3/2}$ level exists close to the two-photon frequency $\omega_{fg}/2$, the behavior of 5S-5D TPTs is much different from that of the nonresonant TPT. Based on a framework previously developed in Ref. [20, 21], the total contribution of resonant TPT interacting with femtosecond pulses via one intermediate state $|i\rangle$ can be given by

$$a_{gf} \propto -\frac{\mu_{fi}\mu_{ig}}{i\hbar^2}[i\pi E(\omega_{ig})E(\omega_{fg}-\omega_{ig}) + \wp\int_{-\infty}^{\infty}\frac{E(\omega)E(\omega_{fg}-\omega)}{\omega_{ig}-\omega}d\omega], \quad (1)$$

where $\wp$ is the principal value of Cauchy, $E(\omega)$ is the electric field magnitude and $\omega_{ig(fg)}$ is ground to intermediate (final) state transition frequency, $\mu_{fi(ig)}$ are the dipole moments from the ground to intermediate (intermediate to final) states. The first term in Eq. (1) is the contribution of S-TPT, whereas the second term is the contribution of D-TPT. As we can see, the D-TPT has a different contribution below and above the $\omega_{ig}$. The contribution from spectral components below the $\omega_{ig}$ is positive while that from above is negative. Taking account of the repetition rate $f_r$ and offset frequency $f_0$ of the femtosecond comb, the second term D-TPT can be given by

$$a_{D-TPT} \propto \sum_m\sum_n |E_m||E_n| \times \left[\frac{1}{\omega_{gi}-2\pi(nf_r+f_0)}+\frac{1}{\omega_{gi}-2\pi(mf_r+f_0)}\right], \quad (2)$$

assuming that the sum of frequencies of m[th] mode and n[th] mode is equal to $\omega_{fg}$. When the [87]Rb atoms are excited by transform-limited pulses, centered on $\omega_{fg}/2$ with broad optical spectrum, the D-TPT integrates over both negative and positive contributions and becomes negligible. Therefore, transform-limited pulses are not optimal for resonant TPT. Similar to the experiment performed in Ref. [20], the first experiment in this work is to prevent the destructive interference by blocking all red (or blue) detuned comb modes.

Figure 1b shows the schematic of the experimental setup. The laser source is an Er-doped fiber-based femtosecond comb operating at ~ 1.5 μm. The output pulse train from the 1.5-μm fiber-based femtosecond comb is firstly amplified and then frequency doubled by a periodically poled lithium niobate (PPLN) waveguide. The generated 780-nm frequency comb has ~ 30-mW average power, with a central wavelength of 776.5nm and a full width at half-maximum bandwidth (FWHM) of approximately 10 nm. A standard 2f-2f configuration pulse shaper with a 1200 grooves/mm grating, a lens and a mirror is used. The grating disperses the ultrashort pulses in spectral domain, so that different spectral teeth are spatially separated. The spatially separated optical spectrum can be manipulated by blocking or choosing the different comb modes. The [87]Rb atoms cell is placed in an oven, and its temperature is stabilized to 70°C. The fluorescence at 420 nm by spontaneous decay from 6P state to 5S is collected using a Fresnel lens and a 420 nm filter to avoid the influence of scattered light from excitation pulses and background light. The TPT signal is detected by a photomultiplier tube (PMT), and recorded by a digital oscilloscope. The [87]Rb cell is placed in a specially designed μ-metal magnetic shield for reducing the influence of external fields.

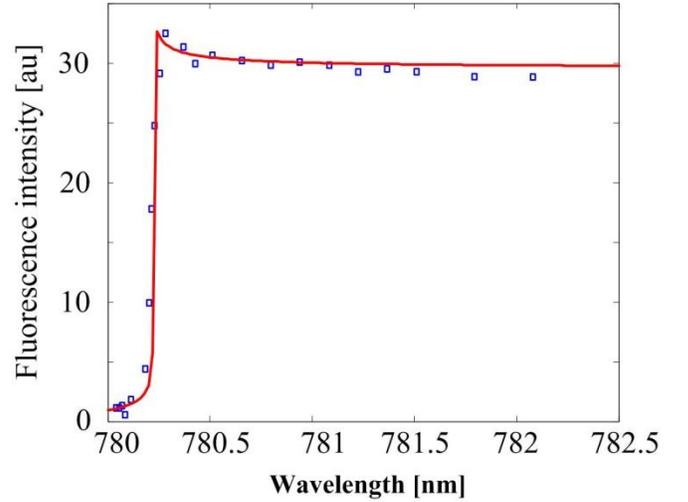

FIG. 2 (color online). The normalized experimental fluorescece intensity (diamonds) and calculated (line) results based on the optical spectrum used in the experiment as a function of the cutoff wavelength.

For comparison with the previous work reported in Ref. [20], the first experiment in this work is performed by blocking parts of the optical spectrum of the excitation pulses. As shown in Fig. 1b, this is achieved by placing a removable shield to block spectral bands symmetrically around $\omega_{fg}/2$. The experimental results (diamonds) are shown in Fig. 2. In Fig. 2, the TPT signal is plotted as a function of the cutoff wavelength together with the numerical simulation calculated by Eq. (1) (the solid line). As the cutoff wavelengths approaching the resonant wavelength, we indeed observe an enhancement of signal amplitude. Keeping the shield closing further, the fluorescence signal rapidly decreased. This TPT changing trend agrees well with that in Ref. [20]. However, only six percent or less of enhancement is achieved in this experiment. The enhancement didn't reach the same level as that in Ref. [20]. This is because the spectrum of the frequency-doubled Er-doped fiber frequency comb is not centered on the two-photon transition frequency $\omega_{fg}/2$, and the spectral bandwidth is not as wide enough as the Ti: sapphire laser used in Ref. [20], so that the amount of positive TPT contribution from the red detuned comb modes is much smaller than that of negative contribution from the blue detuned ones. Therefore this method is effective limited for optical frequency combs with a broad spectrum.

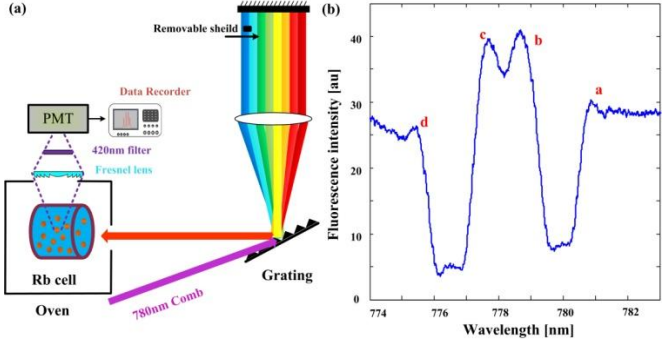

FIG. 3 (color online). (a) General schematic of Experimental setup. A small shield (black square) is used to block part of frequency comb teeth. (b) Experimental results for enhancement of resonant TPT in $^{87}$Rb by scanning the 1.2-nm shield through the whole spectrum of the excitation pulse.

The second experiment is performed to find a more simple and universal quantum coherent control to enhance the resonant TPT rate for optical frequency combs with narrow spectrum bandwidth or not centered at $\omega_{fg}/2$. The experimental setup is shown in Fig. 3a. The laser source used here is identical to the above first experiment. The only difference between these two experiments is the different means of tailoring the comb spectrum. In this experiment, instead of continuously and symmetrically blocking all the red-detuned photons, we remove some certain part of comb modes by utilizing small and removable shields. As shown in Fig. 3a, the shield (the black square) is smaller than the spatial bandwidth of the frequency comb spectrum dispersed by 2f-2f configuration pulse shaper. In this experiment, one shield with about 1.2-mm length, corresponding to 1.2-nm spectrum width, is used to block part of frequency comb teeth to explore the enhancement of TPT rate. Our experimental results are presented in Fig. 3b. These results demonstrate a surprising enhancement of TPT rate. There are four TPT peaks (marked as a, b, c, and d) in Fig. 3b. Maximum enhancements were achieved when the blocking shield was placed between $\omega_{fg}/2$ and the frequencies of the resonant transition. Compared with that of first experiment, the enhancement is greatly improved, reaching a factor of 1.75. Therefore, this quantum coherent control method is much simple with a decent effectiveness.

In the following sections, we present a comprehensive analysis and numerical simulations of this physical process, based on the theory proposed in Ref. [20, 21]. The numerical results are calculated by Eq. (1) and (2), shown in Fig. 4. The dashed line in Fig. 4a represents a 10-nm bandwidth optical spectrum centering at 776.5 nm and obeying a Gaussian distribution, used in the numerical simulations, while the solid line in Fig. 4a represents the D-TPT contribution calculated by Eq. (2). The inset in Fig. 4a shows the detailed information of D-TPT around the resonant wavelength 780.2nm. As the removable gray shield (shown in the inset) is approaching the resonant wavelength 780.2nm from the far red-detuned wavelengths, the corresponding positive contribution is eliminated and the negative contribution from the other side symmetric about $\omega_{fg}/2$ is also removed. From the view of Eq. (2), we can conclude that the blue-detuned comb modes are more far from 780.2nm so that the decrease of positive contributions is larger than that of negative ones. Therefore the total D-TPT signal mainly depends on the negative contribution. When the shield is close to the resonant wavelength, the maximum of positive contribution is blocked; thereby a TPT peak ("a" peak in Fig. 3b) will appear in this place. Along with the shield passing through the resonant frequency, the S-TPT signal is eliminated. As shown in Fig. 3b, we observe a steep decline of the fluorescence signal. Keeping the shield moving away from the resonant frequency, the maximum of the negative contribution is removed whereas the maximum of positive is reserved; hence the "b" peak in Fig. 3b appears in this situation. Because the optical spectrum is blue-detuned about $\omega_{fg}/2$, the "b" peak is much higher than "a" peak. Symmetrically, when the comb modes around the 776nm are blocked by the shield, there will be also two peaks and a deep decline as shown in Fig. 3b.

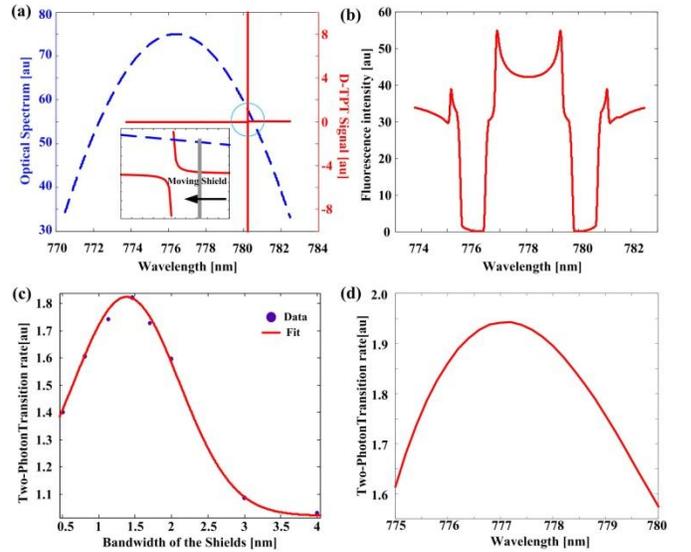

FIG. 4 (color online). (a) A Gaussian optical spectrum (dashed line) of excitation pulses used in the numerical simulation, centering at 776.5 nm; the solid line represents the D-TPT contribution. The inset gives detailed information of D-TPT around the resonant frequency. (b) Calculated simulation results for enhancement of resonant TPT in $^{87}$Rb by scanning a 1.2-nm shield window over the spectrum of the pulse. (c) Normalized TPT rate as a function of the bandwidth of shields. (d) Normalized TPT rate as a function of central wavelengths of the excitation frequency combs with same spectral shape and FWHM.

Figure 4b shows the numerical calculation simulation of the whole aforementioned physical process. In the calculation, we use the optical spectrum shown in Fig. 4a and 1.2-nm spectral bandwidth shield as the numerical conditions. Compared with the experimental results in Fig. 3b, as we can see, a good agreement with the experiment is obtained, although the peaks in the experimental results appear wider than the calculated ones. We attribute the larger widths of the experimental peaks mainly to the linewidth of the comb teeth of the frequency comb, the jitter of the offset frequency, and noise contributions from the electrical system. Furthermore, to find the maximum enhancement, we do more experiments by using different sizes of the removable shields. Figure 4c shows the experimental enhancement (circles) of the TPT rate as a function of sizes of the removable shields. We observe that the maximum enhancement of TPT rate can be achieved by a factor of more than 80%, by using a ~1.5-nm spectral bandwidth shield. To test the universality of this physical mechanism, we perform another numerical calculation as a function of the central wavelengths of excitation frequency combs with same spectral shape and FWHM. The simulation result is presented in Fig. 4d. From 775 nm to 780 nm, more than 50% (maximum 90%) enhancement can be achieved by this quantum coherent control. We can conclude that the mechanism demonstrated here is much effective and universal.

In conclusion, we have shown that resonant two-photon transition rate can be significantly enhanced by simply removing some certain teeth around resonance frequency. This physical mechanism with a particularly simple structure is much more effective and universal for frequency combs with different spectrum bandwidth and central wavelengths, especially for the nonlinear interactions between more and more widely used fiber-based frequency combs and matters. The demonstrated results agree well with the theory. Since this enhancement achieved here is based on a general theory of multiphoton transitions with resonant state, we believe that this mechanism will have a considerable impact on other nonlinear interactions between femtosecond combs and matters [25, 26]. Besides, this mechanism is also particularly meaningful for the Doppler-free TPT DFCS in cold atom system, in which the atomic density is much smaller than that in thermal atom systems [6, 21].

The authors thank the fruitful discussions with Dr. Jiutao Wu from Nanjing Research Institute of Electronics Technology, China. Financial supports of this research by the National Natural Science Foundations of China under Grant No. 61535001 and No. 61371074.

*zhaojianye@pku.edu.cn